\documentclass[%
 reprint,
superscriptaddress,
nofootinbib,
 amsmath,amssymb,
 aps,
]{revtex4-1}

\usepackage{graphicx}
\graphicspath{ {./Figures/} }
\usepackage{dcolumn}
\usepackage{bm}
\usepackage{tablefootnote} 
\usepackage{threeparttable} 
\usepackage{physics}
\usepackage{xcolor}
\usepackage{soul}

\begin{document}


\title{Electron-phonon scattering and thermoelectric transport in $p$-type PbTe\\ from first principles}

\author{Ransell D'Souza}
\email{ransell.dsouza@tyndall.ie}
\affiliation{%
Tyndall National Institute, Lee Maltings, Dyke Parade, Cork T12 R5CP, Ireland 
}%
\author{Jiang Cao}%
\affiliation{%
Tyndall National Institute, Lee Maltings, Dyke Parade, Cork T12 R5CP, Ireland 
}%
\affiliation{%
School of Electronic and Optical Engineering, Nanjing University of Science and Technology, Nanjing 210094, China 
}%
\author{Jos\'{e} D. Querales-Flores}
\affiliation{%
Tyndall National Institute, Lee Maltings, Dyke Parade, Cork T12 R5CP, Ireland 
}%
\author{Stephen Fahy}
\affiliation{%
Tyndall National Institute, Lee Maltings, Dyke Parade, Cork T12 R5CP, Ireland 
}%
\affiliation{%
Department of Physics, University College Cork, College Road, Cork T12 K8AF, Ireland
}%
\author{Ivana Savi\'c}
\email{ivana.savic@tyndall.ie}
\affiliation{%
Tyndall National Institute, Lee Maltings, Dyke Parade, Cork T12 R5CP, Ireland 
}%



\date{\today}

\begin{abstract}
We present a first principles based model of electron-phonon scattering mechanisms and thermoelectric transport at the L and $\Sigma$ valleys in ​$p$-type PbTe, accounting for their thermally induced shifts. Our calculated values of all thermoelectric transport parameters at room temperature are in very good agreement with experiments for a wide range of doping concentrations. Scattering due to longitudinal optical phonons is the main scattering mechanism in $p$-type PbTe, while scattering due to transverse optical modes is the weakest. The L valleys contribute most to thermoelectric transport at 300 K due to the sizeable energy difference between the L and $\Sigma$ valleys. We show that both scattering between the L and $\Sigma$ valleys and additional transport channels of the $\Sigma$ valleys are beneficial for the overall thermoelectric performance of $p$-type PbTe at 300 K. Our findings thus support the idea that materials with high valley degeneracy may be good thermoelectrics.  

\end{abstract}

\maketitle


\section{\label{sec:int}Introduction}
Thermoelectric (TE) materials are capable of converting waste heat into electricity and {\it vice versa}.
The efficiency of the energy conversion in a TE material is defined by the thermoelectric figure of merit $ZT = \frac{\sigma S^2 T}{\kappa_L + \kappa_e}$ \cite{snyder08}, where $\sigma$, $S$, $T$, $\kappa_e$ and $\kappa_L$ are the electrical conductivity, Seebeck coefficient, temperature, electrical thermal conductivity and lattice thermal conductivity, respectively. The figure of merit of a thermoelectric material can be enhanced by either increasing the power factor, $\sigma S^2$, or lowering the total thermal conductivity, $\kappa_L + \kappa_e$.

It has been recently proposed that high valley degeneracy in a material could lead to a large power factor and $ZT$ \cite{pei2011}. Recent experiments argued that tuning the energy levels to achieve high valley degeneracy increases the $ZT$ of a number of materials~\cite{pei2011,liu12,tang15,zhao13,banik15,zheng18,liu18,wang16,kim17}. For example, PbTe and its alloys with other {\rm IV-VI} materials, like PbSe, have energetically close valence band maxima located at the L points and along the $\Sigma$ directions in the Brillouin zone (BZ) \cite{pei2011}. At critical compositions or temperatures where these valleys are aligned, a simultaneous increase in both $\sigma$ and $S$ has been reported for PbTe$_{1-x}$Se$_x$ \cite{pei2011}. This result is surprising, since for most semiconductors, an increase in the conductivity typically reduces the Seebeck coefficient and {\it vice versa} \cite{snyder08}. Moreover, scattering processes between different valleys can reduce carrier lifetimes and electrical conductivity \cite{snyder08}. On the other hand, high valley degeneracy should lead to an increased density of states and enhanced Seebeck coefficient \cite{snyder08}. It is therefore still unclear under which conditions this so-called ``valley convergence'' strategy improves the power factor.

In this work, we establish a theoretical framework to study the temperature driven valley convergence effects on the power factor and figure of merit of $p$-type PbTe~\cite{pei11}. Our recent first principles calculations have predicted that the L and $\Sigma$ valleys become degenerate in energy around 620 K \cite{jose19}, in agreement with a previous {\it ab initio} molecular dynamics simulation \cite{gibbs13}. To understand how this valley convergence affects the thermoelectric properties of PbTe, it is necessary to develop an accurate thermoelectric transport model that can account for the temperature dependence of the electronic band structure.  

Recently, substantial progress has been made in modelling of electronic and thermoelectric transport properties of bulk materials \cite{sohier14,qiu15,li15,fiorentini16,zhou16,gunst16,liu17,ponce18}, where carrier lifetimes due to electron-phonon (el-ph) coupling are calculated from first principles, without any fitting parameters \cite{giustino07,sjakste15,verdi15,giustino17}. However, band structure variations due to temperature are typically not taken into account in these approaches. On the other hand, a theoretical framework to calculate the temperature dependence of electronic states due to el-ph interactions has also been developed \cite{allen76,allen81,allen83} and implemented in first principles codes \cite{gonze11,ponce15}. We have recently combined these advancements and developed a first principles based model of thermoelectric transport in $n$-type PbTe where the temperature induced changes of the band structure were included \cite{jiang18,jiang19}.

In this paper, we develop a first principles based model of thermoelectric transport in $p$-type PbTe that can account for the temperature driven changes of the relative positions of the nearly degenerate L and $\Sigma$ valleys. This model is significantly more complex than that for $n$-type PbTe, where electronic conduction occurs only in the L valleys. To accurately describe acoustic and non-polar optical phonon scattering within the $\Sigma$ valleys, we generalize the Herring and Vogt deformation potential approach \cite{herring56}. We also explicitly account for the intervalley scattering between the L and $\Sigma$ valleys, and among the twelve degenerate $\Sigma$ valleys. All the relevant parameters are calculated from first principles. We also compute the room temperature thermoelectric transport parameters of $p$-type PbTe, which agree very well with available measurements.
As in $n$-type PbTe, longitudinal (transverse) optical phonon scattering is the strongest (weakest) scattering mechanism \cite{jiang18}. The L valleys, which are higher in energy than the $\Sigma$ valleys, dominate thermoelectric transport at 300 K. Even though scattering between the L and $\Sigma$ valleys decreases the electrical conductivity of the L valleys, their Seebeck coefficient is increased and electrical thermal conductivity is decreased.
Consequently, the power factor and figure of merit of the L valleys are improved due to the L-$\Sigma$ intervalley scattering.
The electronic states at the $\Sigma$ valleys further increase all the thermoelectric transport properties, especially at high doping concentrations. As a result, the total figure of merit of $p$-type PbTe is increased at 300 K due to the presence of the $\Sigma$ valleys.

\section{Methodology}
\subsection{Electronic band structure of $p$-type PbTe at 0 K\label{met:eb}}
\begin{figure}[!htbp]
\begin{centering}
\includegraphics[keepaspectratio, width=0.48\textwidth]{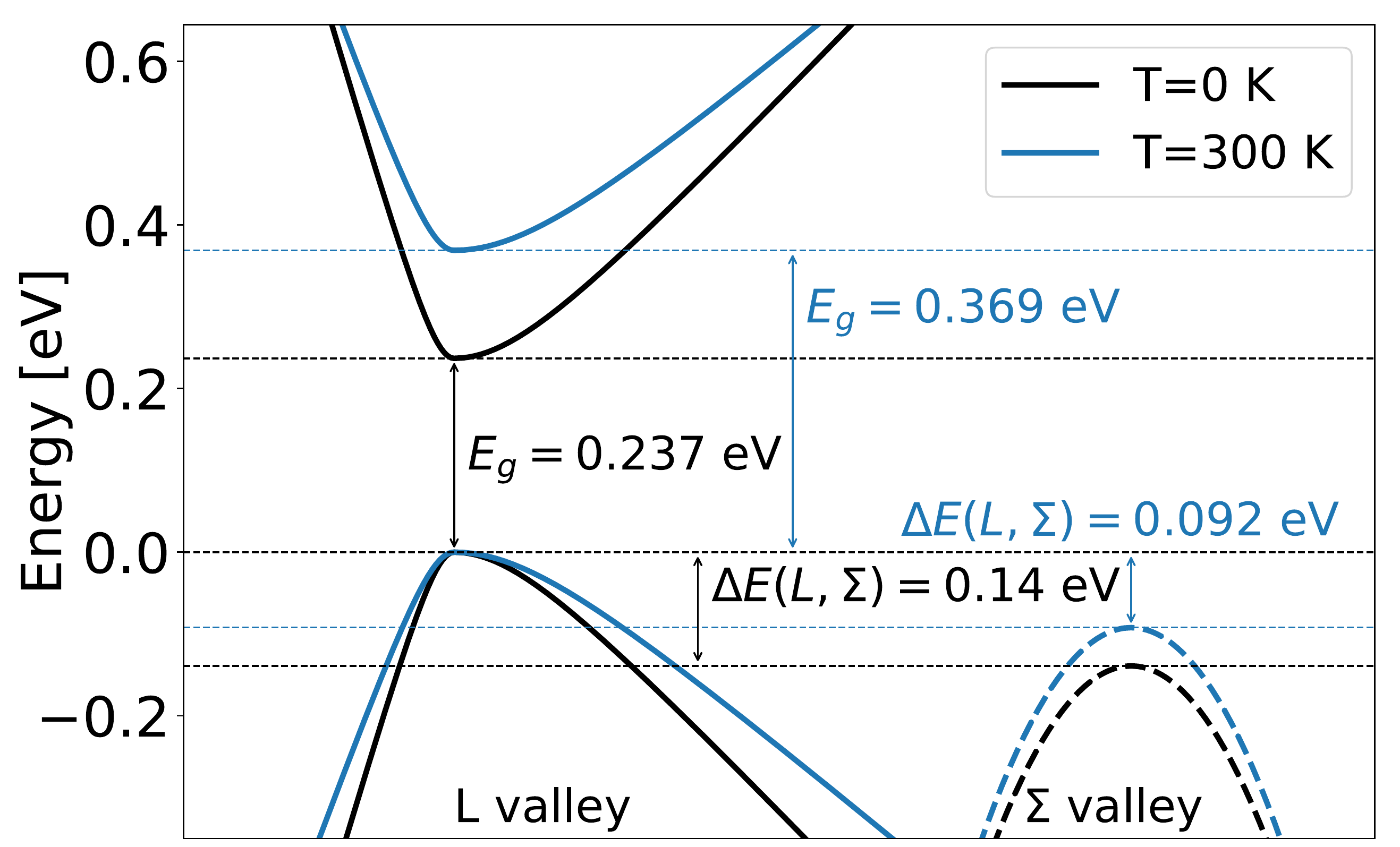}
\end{centering}
\caption{\label{fig:sc_bs} Schematic electronic band structure of PbTe, illustrating the conduction and valence band states at L and valence band states at $\Sigma$ at $0$~K (black) and $300$~K (blue). The horizontal lines represent the energies of the valence band maxima and the conduction band minimum. The states at the L ($\Sigma$) valleys are shown in solid (dashed) lines.}
\end{figure}

The direct narrow band gap in PbTe is located at four equivalent L points in the first BZ, see Fig.~\ref{fig:sc_bs}. In addition to the L valleys, the $\Sigma$ valleys are also populated by holes at higher doping concentrations and temperatures. Twelve equivalent valence band maxima are found midway along the [110] and equivalent $\Sigma$ directions in the first BZ. The energy dispersion of the valence and conduction bands at an L valley and the valence band at a $\Sigma$ valley (three band model) can be characterized with the Kane model derived from the $\bm{k\cdot p}$ Hamiltonian \cite{kane57}
\begin{eqnarray}\label{eq:kane_L}
\frac{\hbar^2}{2}\bigg[\frac{(k^L_{\parallel})^2}{m^L_\parallel} + \frac{(k^L_{\perp})^2}{m^L_\perp} \bigg] = \gamma_{L}(E), 
\end{eqnarray}
\begin{eqnarray}\label{eq:kane_sig}
\frac{\hbar^2}{2}\bigg[\frac{(k^{\Sigma}_{\parallel})^2}{m^{\Sigma}_\parallel} + \frac{(k^{\Sigma}_{\perp_{xy}})^2}{m^{\Sigma}_{\perp_{xy}}} + \frac{(k^{\Sigma}_{\perp_{z}})^2}{m^{\Sigma}_{\perp_{z}}} \bigg] = \gamma_{\Sigma}(E) ,
\end{eqnarray}
\begin{eqnarray}\label{eq:kane}
\gamma_{L(\Sigma)} = E(1+\alpha_{L(\Sigma)}E).
\end{eqnarray}
Eqs.~\eqref{eq:kane_L} and \eqref{eq:kane_sig} describe the band dispersions at the L and $\Sigma$ valleys, respectively. $E$ is the energy of the states with respect to the relevant energy extremum and $\hbar$ is the reduced Planck constant. In Eq.~\eqref{eq:kane_L}, $m^L_{\parallel}$ and $m^L_{\perp}$ are the effective masses for an L valley along the parallel (L-$\Gamma$) and perpendicular (L-W) directions with the wave vector components $k^L_{\parallel}$ and $k^L_{\perp}$ with respect to the L point, respectively. In Eq.~\eqref{eq:kane_sig}, $m^{\Sigma}_{\parallel}$ is the effective mass of a $\Sigma$ valley in the parallel ($\Sigma$-$\Gamma$) direction with the wave vector component $k^{\Sigma}_{\parallel}$ with respect to the $\Sigma$ valley maximum. The effective masses of a $\Sigma$ valley that correspond to the perpendicular directions, [0,0,1] and [1,-1,0], with the wave vector components $k^{\Sigma}_{\perp_{z}}$ and $k^{\Sigma}_{\perp_{xy}}$  with respect to the $\Sigma$ valley maximum, are denoted as $m^{\Sigma}_{\perp_{z}}$ and $m^{\Sigma}_{\perp_{xy}}$, respectively. $\alpha_{L}$ ($\alpha_{\Sigma}$) is the non-parabolicity factor for an L ($\Sigma$) valley. If the coupling of the top valence and bottom conduction bands with all other bands is small, then $\alpha_{L(\Sigma)}=1/E_g^{L(\Sigma)}$, where $E_g^{L(\Sigma)}$ are the direct band gaps at L and $\Sigma$~\cite{kane57}. In the Kane model, we treat the conduction band states at the L valleys as the mirror images in energy of the valence band states near L with respect to the middle of the gap. In Appendix \ref{apdx:kane}, we give the expressions used to compute all the required quantities for the calculation of the scattering rates and TE transport parameters within the Kane model (density of states, group velocities and overlap integrals).

The band parameters for the Kane model, the band gaps at L and $\Sigma$, and the energy difference between the valence band maxima at L and $\Sigma$ were computed using density functional theory (DFT) with the {\sc Quantum espresso} code \cite{qe17,qe09}. In our fit of the DFT electronic band structure with the Kane model, we use $\alpha_{L(\Sigma)}=1/E_g^{L(\Sigma)}$. The DFT calculations were done on a $10\times 10 \times 10$ reciprocal space $\bm k$ grid using the local density approximation (LDA) exchange-correlation. The norm conserving LDA pseudopotential without spin-orbit coupling (SOC) for Pb (Te) using 6$s^2$6$p^2$ (5$s^2$5$p^4$) states were generated from the {\sc pslibrary} code \cite{corso14}. The kinetic and charge density energy cut-off employed in our DFT calculations are 90 Ry and 360 Ry, respectively. 

We use the LDA without SOC for the following reasons.
The LDA with SOC gives the wrong character of the states near the direct narrow band gap at L, {\it i.e.}, the valence and conduction bands are inverted resulting in a negative gap \cite{hummer07,svane10,murphy18}. This also leads to an order of magnitude larger values of the effective masses of $p$-type PbTe compared to experiments \cite{hummer07,svane10,murphy18}.    
Furthermore, the top of the valence band is incorrectly located at $\Sigma$, instead of L \cite{murphy18}.
A generalized gradient approximation including SOC also yields poor band gaps and effective masses compared to experiments \cite{murphy18}. In contrast, we showed that the LDA without SOC and the hybrid Heyd-Scuseria-Ernzerhof (HSE03) functional including SOC \cite{heyd03,heyd04} give a correct physical representation of the electronic states near the band edge in PbTe, resulting in very good agreement of their effective masses with experiments \cite{murphy18}. Furthermore,
the degeneracy of the states near the L and $\Sigma$ valleys that are relevant for transport is the same regardless whether SOC is included in the calculation.
However, the band gap at L obtained using the LDA without SOC is overestimated in comparison to experiments, while that computed with the HSE03 functional agrees with experiments \cite{murphy18}. It is important to use an accurate band gap for electronic transport calculations in narrow gap semiconductors, because the minority carriers can have non-negligible impact on the transport properties. 
Therefore, we use our previously computed value with the 
HSE03 functional \cite{murphy18} in our transport model. A similar description of the electronic band structure for $n$-type PbTe gives excellent agreement between the calculated and experimental thermoelectric transport properties for a range of temperatures and doping concentrations \cite{jiang18,jiang19}.

To further justify the use of the LDA excluding SOC, we note that the computed longitudinal optical (LO) phonon frequencies of PbTe are in much better agreement with room temperature inelastic neutron scattering (INS) measurements when SOC is neglected (see Supplemental Material \cite{supp}). We show in Section \ref{ssec:sr} that the electron-phonon scattering rates in $p$-type PbTe are dominated by LO phonons, so they must be accurately described in our calculations of the thermoelectric parameters. Acoustic phonon frequencies are almost identical in the two types of calculations (with and without SOC) and agree with experiments. We have also shown that including SOC only weakly affects acoustic deformations potentials of PbTe \cite{murphy18}. We note that there are differences among the measured soft transverse optical (TO) phonon frequencies near the zone center at 300~K \cite{delaire11,li14,cochran66}. In the INS experiment carried out by Delaire {\it et al.} \cite{delaire11}, there are two peaks in the scattering intensity corresponding to the zone center TO mode as a result of strong anharmonicity. The lower peak at $\Gamma$ is in good agreement with earlier INS measurements by Cochran {\it et al.}~\cite{cochran66} and our computed TO frequency with SOC. However, the upper peak that has the larger scattering intensity \cite{li14} is in good agreement with our TO frequency without SOC \cite{supp}. Nevertheless, the TO modes have negligible effects on the total scattering rates of $p$-type PbTe (Section \ref{ssec:sr}) and taking the TO frequency from the calculations including or excluding SOC would not have any effect on the thermoelectric transport properties.

\subsection{Temperature variation of the electronic band structure}

We have calculated the temperature dependence of the band gap at L and the energy difference between L and $\Sigma$ from first principles, accounting for thermal expansion and el-ph contributions as detailed in Ref.~\cite{jose19}. The el-ph contribution to the temperature renormalization of electronic states is computed using the non-adiabatic Allen-Heine-Cardona approach \cite{allen76,allen81,allen83} and density functional perturbation theory (DFPT)~\cite{gonze97,baroni01}. For temperatures larger than $\sim 50$~K, the variation of the band gap at L with temperature can be considered as linear\footnote{The Debye temperature of PbTe is $\sim 150$~K.} \cite{jose19,jiang19}
\begin{eqnarray}\label{eq:EgT}
E^{L}_g(T) = E^{L}_g(0)+ \frac{\partial E^{L}_g}{\partial T}\times T.
\end{eqnarray}
Here $E^L_g(0)$ is the HSE03 value of the band gap at L. $\frac{\partial E^{L}_g}{\partial T}$ is the temperature coefficients calculated in Ref.~\cite{jose19}, and $T$ is the temperature. The energy difference between the valence band maximum at L and $\Sigma$, $\Delta E_{L,\Sigma}(T)$, is calculated at 0 K using DFT and its 300 K value is taken from Ref.~\cite{jose19}.

To understand our treatment of the effective masses of the valence band states near L, it is important to note that the top valence band state at L is very close in energy to the bottom conduction band state at L (the direct band gap at L is $\sim 0.2$ eV at 0 K), while all other band states at L are much further away in energy (more than 1 eV). This is a classic example of the situation where we have strong coupling between two bands, which is well described using the Kane model. In the derivation of the Kane model, it can be shown that the effective masses are proportional to the band gap \cite{dresselhaus08}. Therefore, we deduce that the effective masses at L scale with temperature as:
\begin{eqnarray}\label{eq:m_Eg}
\frac{m_{\parallel,\perp}^{L}(T)}{m_{\parallel,\perp}^{L}(0)} = \frac{E^{L}_g(T)}{E^{L}_g(0)}.
\end{eqnarray}
We also note that Eq.~\ref{eq:m_Eg}  has been experimentally verified for a number of III-V and II-VI materials \cite{madelung87,yu05}. Since $\alpha_{L}=1/E_g^{L}$, the temperature dependence of the non-parabolicity term at L is given as:
\begin{eqnarray}
\frac{\alpha_{L}(T)}{\alpha_{L}(0)} = \frac{E^{L}_g(0)}{E^{L}_g(T)}.
\end{eqnarray}

To verify the validity of Eq.~\ref{eq:m_Eg}, we have explicitly calculated the temperature dependence of the effective masses at L as done in Ref.~\cite{jose19}. The temperature renormalization of the band structure due to el-ph interaction is computed using the non-adiabatic Allen-Heine-Cardona formalism within DFPT implemented in the {\sc Quantum Espresso} code. Similarly, the temperature renormalization due to thermal expansion is calculated from DFT using the lattice thermal expansion coefficient given in Ref.~\cite{jose19}. Our first principles results show that the Kane model accurately describes the temperature effects on the electronic band structure near L, {\it i.e.}, the effective masses are proportional to the direct band gap at L. The valence band effective masses at L calculated at 300 K using the Kane model and the temperature dependence of the band gap at L obtained from first principles are only 3\% smaller than the corresponding effective masses computed from first principles. Overall, the temperature variations of the band gap and the effective masses at L are large ($\sim 55$\% at 300 K with respect to 0~K).

In contrast, the energy of the top valence band state at $\Sigma$ is relatively far from all other band states at $\Sigma$ (more than 1 eV). Therefore, the top valence band near $\Sigma$ is a classic example of the band that is weakly perturbed by other bands and exhibits nearly parabolic behavior (the non-parabolicity factor used in the fit is only 1/1.35 eV$^{-1}$, where 1.35 eV is the direct band gap at $\Sigma$ at 0 K). In this case, we expect negligible temperature variation of the effective masses at $\Sigma$. To check this assumption, we have explicitly calculated the valence band effective masses at $\Sigma$ due to thermal expansion and el-ph interaction from first principles. We have found only a 4\% increase in the density of states effective mass at 300 K compared to 0~K. This finding justifies the use the 0~K effective masses of the $\Sigma$ valleys in our calculations at 300 K.

The temperature renormalized effective masses at L allow us to obtain the temperature variations of the energy dispersions of the valence and conduction band states at the L valleys given by Eqs.~\eqref{eq:kane_L}-\eqref{eq:kane}. Their relative positions are computed from the temperature dependence of the band gap at L, while their relative positions with respect to the $\Sigma$ valleys are obtained from the temperature dependence of the energy difference between the valence band maxima at L and $\Sigma$. Fig.~\ref{fig:sc_bs} shows a schematic of the three bands at L and $\Sigma$ for $T=0$~K and $T=300$~K. These temperature dependent band dispersions modify the electronic density of states and the group velocities that directly influence the electron-phonon scattering rates and the thermoelectric transport properties, which will be described below.

We do not include the temperature variations of the phonon band structure and electron-phonon matrix elements in the present calculations. This is partly because our computed phonon dispersion at 0 K agrees well with the measured phonon frequencies at 300 K \cite{delaire11, li14, cochran66, supp}. Furthermore, the phonon frequencies of PbTe exhibit weak temperature dependence at temperatures larger than 300~K~\cite{delaire11,cochran66,romero15,ribeiro18}, while the electronic energies change linearly for such temperatures~\cite{jose19}. For example, the TO frequencies at $\Gamma$, which vary most with temperature, change by less than $10$\% between 300 K and 600 K \cite{delaire11,romero15}, while the band gap and effective masses at L increase by $\sim 35$\%. Therefore, temperature affects the relative energies of the electronic states in PbTe more strongly than the phonon frequencies in the temperature range relevant for thermoelectric operation. Moreover, we expect from perturbation theory that electronic energies and phonon frequencies are modified more strongly by external perturbations than electronic wave functions and phonon eigenvectors. Since electron-phonon matrix elements are determined by phonon frequencies and eigenvectors and electronic wave functions, the effect of the temperature dependence of the electron-phonon matrix elements on the scattering rates of PbTe will be smaller than the effect of the temperature dependence of the electronic band structure and can be neglected at first order.

\subsection{Thermoelectric transport\label{sec:trans}}
We calculate thermoelectric transport properties by solving the Boltzmann transport equation using the spherical transformation of the Kane Hamiltonian \cite{ridley99}. For each valley and carrier type (electrons or holes), the electrical conductivity, Seebeck coefficient and electrical thermal conductivity for zero electric field across a material are respectively given as \cite{ridley99,jiang19}
\begin{eqnarray}
\sigma &=& \frac{e^2 m_d^{3/2}}{\pi^2}\int_0^\infty \frac{-\partial f^0}{\partial E}\tau \overline{v}^2k^{*2}dk^*, \label{eq:cond}\\
S &=& \frac{-m_d^{3/2}e}{T\sigma \pi^2} \int_0^\infty \frac{-\partial f^0}{\partial E}\tau \overline{v}^2(E-E_F) k^{*2}dk^*, \label{eq:S}\\
\kappa_0 &=& \frac{m_d^{3/2}}{T \pi^2} \int_0^\infty \frac{-\partial f^0}{\partial E}\tau \overline{v}^2(E-E_F)^2 k^{*2}dk^*, \label{eq:kappa}
\end{eqnarray}
where $f^0$ is the Fermi-Dirac distribution, $e$ is the electronic charge, $k^*$ are the rescaled wave vectors with respect to the valley minimum in the spherical coordinate system, and $E_F$ is the Fermi level. $m_d$ is the density of states (DOS) effective mass, $\overline{v}$ is the average group velocity, and $\tau$ is the carrier relaxation time. The DOS effective masses are given as $m^L_d = [m^L_{\parallel}(m^L_{\perp})^2]^{\frac{1}{3}}$ for an L valley and $m^{\Sigma}_d = (m^{\Sigma}_{\parallel}m^{\Sigma}_{\perp_{xy}}m^{\Sigma}_{\perp_{z}})^{\frac{1}{3}}$ for a $\Sigma$ valley. In the case of holes (electrons), the integration is carried out from a valence band maximum (a conduction band minimum) towards lower (higher) energies. 

The total conductivity in our three band model including the L and $\Sigma$ valleys is calculated using
\begin{eqnarray}\label{eq:sigma}
\sigma_{tot} = N_L\sigma_{h_L} + N_{\Sigma}\sigma_{h_{\Sigma}} + N_L\sigma_{e_L},
\end{eqnarray}
where $N_L = 4$ and $N_{\Sigma} = 12$ are the number of the L and $\Sigma$ valleys, respectively. The indices $h_L$, $h_{\Sigma}$ and $e_L$ denote the contributions to $\sigma$ (and $S$ and $\kappa_e$) from holes at the L valleys, holes at the $\Sigma$ valleys and electrons at the L valleys, respectively. The total Seebeck coefficient in the three band model is \cite{vineis08} 
\begin{eqnarray}
  S_{tot} = \frac{1}{\sigma_{tot}}(N_L S_{L_e} \sigma_{L_e} + N_L S_{L_h} \sigma_{L_h}
  \nonumber \\
  + N_{\Sigma} S_{{\Sigma}_h} \sigma_{{\Sigma}_h} ).
\end{eqnarray}  
The total electrical thermal conductivity for zero electric current across a material is given as
\begin{eqnarray}
\kappa_{e_{tot}} = N_L\kappa_{h_L} + N_{\Sigma}\kappa_{h_{\Sigma}} + N_L\kappa_{e_L} -S_{tot}^2\sigma_{tot}T.
\end{eqnarray}
The total power factor and thermoelectric figure of merit are defined by $\sigma_{tot}S_{tot}^2$ and $\frac{ \sigma_{tot}S_{tot}^2T}{\kappa_{e_{tot}}+\kappa_L}$, respectively.
The lattice thermal conductivity value used to calculate $ZT$ at 300~K is $\kappa_L$=2.4~W/(mK), obtained from our recent first principles calculations~\cite{murphy16}.

\subsection{\label{sec:sr} Scattering rates and relaxation times}
From the first order perturbation theory, the scattering rate due to electron-phonon coupling can be expressed as \cite{ridley99}
\begin{eqnarray}\label{eq:sr}
W_{i} = \sum_f \frac{2 \pi}{\hbar}|\bra{f}H_{ep}\ket{i}|^2\delta(E_f - E_i),
\end{eqnarray}
where $\ket{i}$ and $\ket{f}$ are the initial and final states, respectively, and $E_i$ and $E_f$ are their energies. The delta function represents energy conservation. The sum is over all the final states. Expanding the wave functions of the initial and final states using the products of Bloch and harmonic oscillator wave functions, the electron-phonon matrix elements read~\cite{ridley99}
\begin{equation}\label{eq:el_ph}
\begin{split}
&|\bra{f}H_{ep}\ket{i}|^2 = |M_{\lambda,{\bm q}}|^2  \\
 &= \frac{\hbar}{2m \omega_{\lambda,{\bm q}}}C_{{\lambda},{\bm q}}^2I^2({\bm k},{\bm k'})\delta_{\bm{k'},\bm{k \pm q}}\bigg(n(\omega_{\lambda,{\bm q}})+\frac{1}{2} \mp \frac{1}{2} \bigg).
\end{split}
\end{equation}
$I({\bm k},{\bm k'})$ is the overlap integral between the two Bloch states with the wave vectors ${\bm k}$ and ${\bm k'}$. $C_{\lambda,{\bm q}}$ is the coupling coefficient between these two states via a phonon mode $\lambda$ with the wave vector ${\bm q}$, which satisfies momentum conservation given by the Kronecker delta. $\omega_{\lambda,{\bm q}}$ is the phonon frequency and $n(\omega_{\lambda,{\bm q}})$ is the Bose-Einstein distribution. $m$ is the mass of the atoms in the unit cell. The upper (lower) signs in Eq.~\eqref{eq:el_ph} are for absorption (emission) of a phonon. The scattering rate then yields
\begin{equation}\label{eq:sr2}
W(\bm{k}) = \sum_{\bm{q}} \sum_{\lambda} W_{\lambda}(\bm{k},\bm{k'})\delta_{\bm{k'},\bm{k \pm q}}\delta(E_{\bm k'}-E_{\bm k}\mp \hbar \omega_{\lambda,{\bm q}}),
\end{equation}
where $W_{\lambda}({\bm k},{\bm k'})=2\pi|M_{\lambda,{\bm q}}|^2/\hbar$. The relaxation time is given as~\cite{sohier14,ma18}
\begin{eqnarray}\label{eq:rt}
\tau^{-1}(\bm{k}) &=& \sum_{\bm{q}}\sum_{\lambda}\big(1-\cos\theta_{\bm{k'}}\big) \frac{1-f^0(E_{\bm{k'}})}{1-f^0(E_{\bm{k}})}\nonumber \\ 
&\times& W_{\lambda}(\bm{k},\bm{k'})\delta_{\bm{k'},\bm{k \pm q}}\delta(E_{\bm k'}-E_{\bm k}\mp \hbar \omega_{\lambda,{\bm q}}),
\end{eqnarray}
where $\cos\theta_{\bm{k'}}$ is the angle between $\bm{k}$ and $\bm{k'}$.

More technical details about the scattering rate and relaxation time calculations are given in Appendix~\ref{apdx:tsr}. The contributions to the total relaxation time in our model are due to intervalley and intravalley scattering (including Fr\"ohlich interaction) and ionized impurities. These contributions are added using Matthiessen's rule. What follows next are the details of the intervalley and intravalley acoustic and non-polar optical contributions that are specific for $p$-type PbTe. Fr\"ohlich and ionized-impurity relaxation times, which are the same as in $n$-type PbTe, are given in Ref.~\cite{jiang18}.

\subsection{\label{ssec:ivs}Intervalley and zero order intravalley non-polar optical scattering}
Conservation of momentum imposes that scattering from one valley to another occurs only via long wave vector phonons near the Brillouin zone edge. These phonon frequencies generally have weak wave vector dependence. The coupling term $C_{\lambda,{\bm q}}$ in Eq.~\eqref{eq:el_ph} characterizing the intervalley scattering between valleys $i$ and $j$ due to a phonon with frequency $\omega_{\lambda}$ can be approximated by a constant deformation potential (DP), $\Xi_{ij}^{\lambda}$ \cite{ridley99}.
The corresponding relaxation time is \cite{ridley99}

\begin{eqnarray}
\label{eq:inter}
(\tau^{ij}_\lambda({\bm k}))^{-1}
&=& \frac{\pi V (\Xi_{ij}^{\lambda})^2}{m \omega_{\lambda}} \bigg(n(\omega_{\lambda})D_j(E_{\bm k} - \Delta E_{ij} + \hbar\omega_\lambda)  \nonumber \\
&+& (n(\omega_{\lambda})+1)D_j(E_{\bm k} - \Delta E_{ij} - \hbar\omega_\lambda) \bigg),
\end{eqnarray}
where $V$ is the unit cell volume, $D_j(E)$ is the density of states of valley $j$, and $\Delta E_{ij}$ is the energy difference between the valence band maxima of valleys $i$ and $j$. Scattering between the valence band maxima at L is forbidden by symmetry in PbTe, similarly to that of the conduction band minima at L \cite{jiang18}. This symmetry forbidden scattering stems from the inversion symmetry of PbTe with respect to its constituent atoms \cite{jiang18}. In contrast, the deformation potentials for scattering between the L and $\Sigma$ valleys, and different $\Sigma$ valleys, are not zero. We calculate the values of all these deformation potentials using DFPT (given in Supplemental Material~\cite{supp}).

Eq.~\eqref{eq:inter} is also valid for intravalley scattering ($i=j$) by short wave vector non-polar optical phonons \cite{ridley99}. This type of scattering is also forbidden by symmetry at the L points both for the valence and conduction bands \cite{jiang18}, but it is allowed for the $\Sigma$ valence band maxima (see Supplemental Material~\cite{supp}). Eq.~\eqref{eq:inter} represents the zero order contribution to intravalley non-polar optical scattering. 

\subsection{\label{sec:fos} Intravalley acoustic and first order non-polar optical scattering }
Intravalley scattering is restricted to long wavelength phonon modes due to their energy and momentum conservation. To describe intravalley scattering due to acoustic phonons, we adopt the method described by Fischetti and Laux \cite{fischetti96}, Murphy-Armando and Fahy \cite{armando08} and Murphy {\it et al.} \cite{murphy18}. The interaction between electrons and acoustic phonons in the limit $|{\bm q}|\rightarrow 0$ is characterized by a slowly varying potential, which can be expressed as \cite{herring56,yu05}
\begin{eqnarray}\label{eq:Hep}
H_{ep} = \sum_{\alpha\beta}\Xi_{\alpha\beta}S_{\alpha\beta} ({\bf r}),
\end{eqnarray}
where $\alpha$ and $\beta$ are the Cartesian coordinates, $\Xi$  is the acoustic deformation potential tensor, and $S({\bf r})$ is the local strain tensor at {\bf r}. A second rank tensor has a maximum of six components. However, when the valleys in a cubic material lie on $\Gamma$-L, $\Gamma$-X or $\Gamma$-K high symmetry axes, symmetry constrains dictate that all DP tensor components can be characterized in terms of either two or three independent ones. For an L (or X) valley, the two linearly independent deformation potentials are denoted by $\Xi_d$ and $\Xi_u$. $\Xi_d$ describes the shift because of the dilation in two directions perpendicular to the valley axis \cite{herring56}. $\Xi_u$ represents the shift due to the uniaxial shear associated with an elongation along the valley axis and a contraction in two perpendicular directions \cite{herring56}. For a $\Sigma$ valley, an additional deformation potential $\Xi_p$ characterizes the shift due to sheer in the [001] plane which includes the valley axis \cite{herring56}. The symmetry restrictions on the six components of the tensor are tabulated in Table \ref{tab:sym_res} for L and $\Sigma$ valleys \cite{herring56}.
\begin{center}
\begin{table}[h]
\caption{\label{tab:sym_res}Expressions for the six components of the deformation potential tensor due to the cubic symmetry restrictions for L and $\Sigma$ valleys.}
\renewcommand{\arraystretch}{1.4}
\begin{tabular}{ c | c | c }
Valley & L & $\Sigma$ \\
\hline
Valley direction & [111] & [110]\\
\hline
\hline
$\Xi_{1}$ = $\Xi_{xx}$ & $\Xi_d$ + $\frac{\Xi_u}{3}$ & $\Xi_d$ + $\Xi_u$ - $\frac{\Xi_p}{2}$ \\ 
$\Xi_{2}$ = $\Xi_{yy}$ & $\Xi_d$ + $\frac{\Xi_u}{3}$ & $\Xi_d$ + $\Xi_u$ - $\frac{\Xi_p}{2}$ \\ 
$\Xi_{3}$ = $\Xi_{zz}$ & $\Xi_d$ + $\frac{\Xi_u}{3}$ & $\Xi_d$ - $\Xi_u$ +  $\Xi_p$ \\ 
$\Xi_{4}$ = $\Xi_{yz}$ = $\Xi_{zy}$ & $\frac{\Xi_u}{3}$ & 0 \\ 
$\Xi_{5}$ = $\Xi_{xz}$ = $\Xi_{zx}$ & $\frac{\Xi_u}{3}$ & 0 \\ 
$\Xi_{6}$ = $\Xi_{xy}$ = $\Xi_{yx}$ & $\frac{\Xi_u}{3}$ & $\frac{\Xi_p}{2}$\\ 
\end{tabular}
\end{table}
\end{center}

We obtain electron-phonon matrix elements due to intravalley acoustic scattering using the deformation potential tensor described above. Since the acoustic phonon frequencies near the zone center are small, we use $n(\omega_{\lambda,{\bm q}}) \approx \frac{k_BT}{\hbar \omega_{\lambda,{\bm q}}}$, where $k_B$ is the Boltzmann constant, $\lambda=1,2,3$ and $|{\bm q}|\rightarrow 0$. At room temperature, the equipartition between absorption and emission processes holds, {\it i.e.} $n(\omega_{\lambda,{\bm q}}) \gg 1/2$. Assuming the linear phonon dispersion in the long wavelength limit, the contributions from longitudinal (L) and transverse (T) acoustic branches to the electron-phonon matrix element are given as \cite{herring56,ridley99}
\begin{eqnarray}\label{eq:bc} 
|M_{L(T)}^{ac}|^2 = \frac{k_B T I^2({\bm{k,k'}})}{V}\frac{\sum_{ij}\Xi_i \Xi_j f_i f_j}{\sum_{ij} c_{ij} f_i f_j}.
\end{eqnarray}
Here $c_{ij}$ are the components of the 6 $\times$ 6 elastic constant matrix, where $i$ and $j$ denote the reduced indices as defined in Table \ref{tab:sym_res}. $f$ is the 3$\times$3 polarization tensor determined from
\begin{eqnarray}\label{eq:pol}
\begin{gathered}
f_i = a_iq_i\ ;\ i = 1,2,3,\\
f_4 = a_2q_3 + a_3q_2, \\
f_5 = a_1q_3 + a_3q_1, \\
f_6 = a_1q_2 + a_2q_1,
\end{gathered}
\end{eqnarray}
where ${\bm a}$ is the atomic displacement for an acoustic phonon with wave vector ${\bm q}$. We calculate the effective elastic constants and polarization vectors by solving the generalized equation of motion for the atomic displacement \cite{ridley99}
\begin{widetext}
\begin{align}\label{mat:we}
\begin{pmatrix}
c_{44} + \alpha^2(c_{12}+c_{44}+c^{*}) - \rho v_s^2 & (c_{12}+c_{44})\alpha\beta & (c_{12}+c_{44})\alpha\gamma \\
(c_{12}+c_{44})\alpha\beta & c_{44} + \beta^2(c_{12}+c_{44}+c^{*}) - \rho v_s^2 & (c_{12}+c_{44})\beta\gamma \\
(c_{12}+c_{44})\alpha\gamma & (c_{12}+c_{44})\beta\gamma & c_{44} + \gamma^2(c_{12}+c_{44}+c^{*}) - \rho v_s^2
\end{pmatrix}
\begin{pmatrix}
a_1 \\
a_2 \\
a_3
\end{pmatrix}
= 0,
\end{align}
\end{widetext}
\noindent where $\rho$ is the mass density, $v_s$ is the speed of sound and $c^* = c_{11} - c_{12} -2c_{44}$. The direction cosines of the wave vector $\bm q$ are denoted by $\alpha$, $\beta$ and $\gamma$. Herring and Vogt \cite{herring56} have given the expressions for the electron-phonon matrix elements due to acoustic modes along the high symmetry directions for L and $\Delta$ valleys. We can obtain the matrix elements for longitudinal and transverse acoustic branches along any ${\bm q}$ direction by substituting the polarization components given by Eqs.~\eqref{eq:pol} and \eqref{mat:we} and the DP components from Table~\ref{tab:sym_res} in Eq.~\eqref{eq:bc}. The acoustic el-ph matrix elements for L and $\Sigma$ valleys along the high symmetry directions are listed in Table~\ref{tab:LaTa-cont}. We reproduce Herring and Vogt results for an L valley \cite{herring56}, and give the equivalent results for a $\Sigma$ valley.

\begin{table}
\caption{\label{tab:LaTa-cont} Longitudinal acoustic (LA) and transverse acoustic (TA) mode contributions to the square of the electron-phonon matrix elements within L and $\Sigma$ valleys. $\Xi$ and $c$ are the 6$\times$6 deformation potential and elastic constant tensors, respectively. $c^*$ is the measure of elastic anisotropy, $c^* = c_{11} - c_{12} -2c_{44}$. ${\bm q}$ represents the phonon wave vector.}
\resizebox{0.45\textwidth}{!}{
\renewcommand{\arraystretch}{2.2}
  \begin{tabular}{ c | c | c | c | c }
\hline
  $\Sigma$ valley & ${\bm q}$ direction & LA & TA$_1$ & TA$_2$ \\
  \hline \hline
[110] & [110] & $\frac{(\Xi_d + \Xi_u)^2}{\frac{c^*}{2}+c_{12}+2c_{44}}$ & 0 & 0 \\
 & [100] & $\frac{(\Xi_d + \Xi_u - \frac{\Xi_p}{2})^2}{  c^*+c_{12}+2c_{44}  }$ & $\frac{\frac{1}{4}\Xi_p^2}{c_{44}}$ & 0 \\
 & [111] & $\frac{(\Xi_d + \frac{\Xi_u}{3} + \frac{\Xi_p}{3})^2}{  \frac{c^*}{3} + c_{12} + 2c_{44}}$ & 0 & $\frac{\frac{2}{9}(\Xi_p - 2\Xi_u)^2}{c_{44} + \frac{c^*}{3} }$ \\
 & [1$\overline{1}0$] & $\frac{(\Xi_d + \Xi_u - \Xi_p)^2}{\frac{c^*}{2} + c_{12} + 2c_{44}}$ & 0 & 0 \\
 & [1$\overline{1}$1] & $\frac{(\Xi_d + \frac{\Xi_u}{3} - \frac{\Xi_p}{3})^2}{  \frac{c^*}{3} + c_{12} + 2c_{44}}$ & 0 &  $\frac{\frac{8}{9}(\Xi_p - \Xi_u)^2}{c_{44} + \frac{c^*}{3} }$ \vspace{0.05em}\\
\hline
   L valley & ${\bm q}$ direction & LA & TA$_1$ & TA$_2$ \\
  \hline \hline
[111] & [110] & $\frac{(\Xi_d + \frac{2\Xi_u}{3})^2}{\frac{c^*}{2}+c_{12}+2c_{44}}$ & $\frac{\frac{2}{9}\Xi_u^2}{c_{44}}$ & 0 \\
 & [100] & $\frac{(\Xi_d + \frac{\Xi_u}{3})^2}{  c^*+c_{12}+2c_{44}  }$ & $\frac{\frac{1}{9}\Xi_u^2}{c_{44}}$ & $\frac{\frac{1}{9}\Xi_u^2}{c_{44}}$ \\
 & [111] & $\frac{(\Xi_d + \Xi_u)^2}{ \frac{c^*}{3} + c_{12} + 2c_{44}}$ & 0 & 0 \\
 & [1$\overline{1}0$] & $\frac{\Xi_d^2}{\frac{c^*}{2} + c_{12} + 2c_{44}}$ & 0 & 0 \\
 & [1$\overline{1}$1] & $\frac{(\Xi_d + \frac{\Xi_u}{9})^2}{  \frac{c^*}{3} + c_{12} + 2c_{44}}$ & $\frac{\frac{2}{27}\Xi_u^2}{c_{44} + \frac{c^*}{3} }$ &  $\frac{\frac{2}{81}\Xi_u^2}{c_{44} + \frac{c^*}{3} }$ \vspace{0.05em}\\
 \hline
\end{tabular}}
\end{table}

Since the zero order intravalley scattering by non-polar optical modes is either forbidden by symmetry (at the L points) or rather weak (at the $\Sigma$ valence band maxima), we also include the first order intervalley scattering by these modes in our model \cite{jiang18,ridley99,harrison56,ferry76}. In this case, the linear $\bm{q}$ dependence of the corresponding el-ph matrix elements is taken into account, similarly as in the case of long wavelength acoustic phonons. The resulting Hamiltonian due to the interaction between electrons and long wavelength optical phonons has the same form as that for acoustic phonons (Eq.~\eqref{eq:Hep}), except that instead of strain we account for optical phonon displacement. The el-ph matrix elements in the limit $|{\bm q}|\rightarrow 0$ are then given as \cite{jiang18,ridley99}
\begin{eqnarray}
\begin{aligned}\label{eq:op}
|M^{op}_{L(T)}|^2 &= \frac{\hbar I^2(\bm{k,k'})}{2m\omega_{L(T)}^{op}}\left(n(\omega_{L(T)}^{op}) + \frac{1}{2} \mp \frac{1}{2}\right) \\
&\times \sum_{i,j}\Xi_i^{op}\Xi_j^{op} f_i^{op} f_j^{op},
\end{aligned}
\end{eqnarray}
where $\omega_{L(T)}^{op}$ are the longitudinal (transverse) optical phonon frequencies, and $\Xi_i^{op}$ and $f_i^{op}$ are the optical deformation potential and polarization tensors, respectively, equivalent to the acoustic ones given in Table~\ref{tab:sym_res} and Eq.~\eqref{eq:pol}. We approximate the longitudinal optical phonon frequency to be a constant and the transverse optical frequency to have a ${\bm q}$ dependence of the form $\omega_{TO}({\bm q}) = \omega^{\Gamma}_{TO} + \frac{\partial^2\omega_{TO}}{\partial |\bm{q}|^2}|\bm{q}|^2$ in the long wavelength limit, which is characteristic for PbTe~\cite{murphy16}.

Further details of our calculations of the el-ph matrix elements due to intravalley acoustic and first order non-polar optical phonons are given in Appendix \ref{apdx:azimuth}. To compute the intravalley deformation potentials for both optical and acoustic phonon modes, we fit the calculated el-ph matrix elements using DFPT (see Eq.~\eqref{eq:me_dfpt} in Appendix \ref{apdx:DPfDFPT}) in the limit $\bm q$ $\rightarrow$ 0 with the equations given in Table \ref{tab:LaTa-cont}, see Appendix \ref{apdx:intra}. The calculated values of the intravalley acoustic and optical deformation potentials for $p$-type PbTe are listed in Table \ref{tab:dp}. The elastic constants and optical phonon frequencies were also obtained using DFPT and are given in our previous work on $n$-type PbTe \cite{jiang18}. 
\begin{table}[h]
\caption{\label{tab:dp} Computed values of the intravalley acoustic and optical deformation potentials for the L and $\Sigma$ valleys of $p$-type PbTe.}
\renewcommand{\arraystretch}{1.2}
\begin{tabular}{ c | c | c | c | c}
Valley & Phonons & $\Xi_d$ (eV) & $\Xi_u$ (eV) & $\Xi_p$ (eV) \\
  \hline \hline
 L & acoustic &-3.22 & 7.42 & - \\
 $\Sigma$ & acoustic & 0.034 & 1.869 & 5.547 \\
 L & optical & 22.66 & -30.87 & - \\
 $\Sigma$ & optical & -22.96 & 13.69 & 19.66 \\
\end{tabular}
\end{table}

\section{Results and discussion}
\subsection{Electronic band structure}
\begin{figure}[!htbp]
\begin{centering}
\includegraphics[keepaspectratio, width=\linewidth]{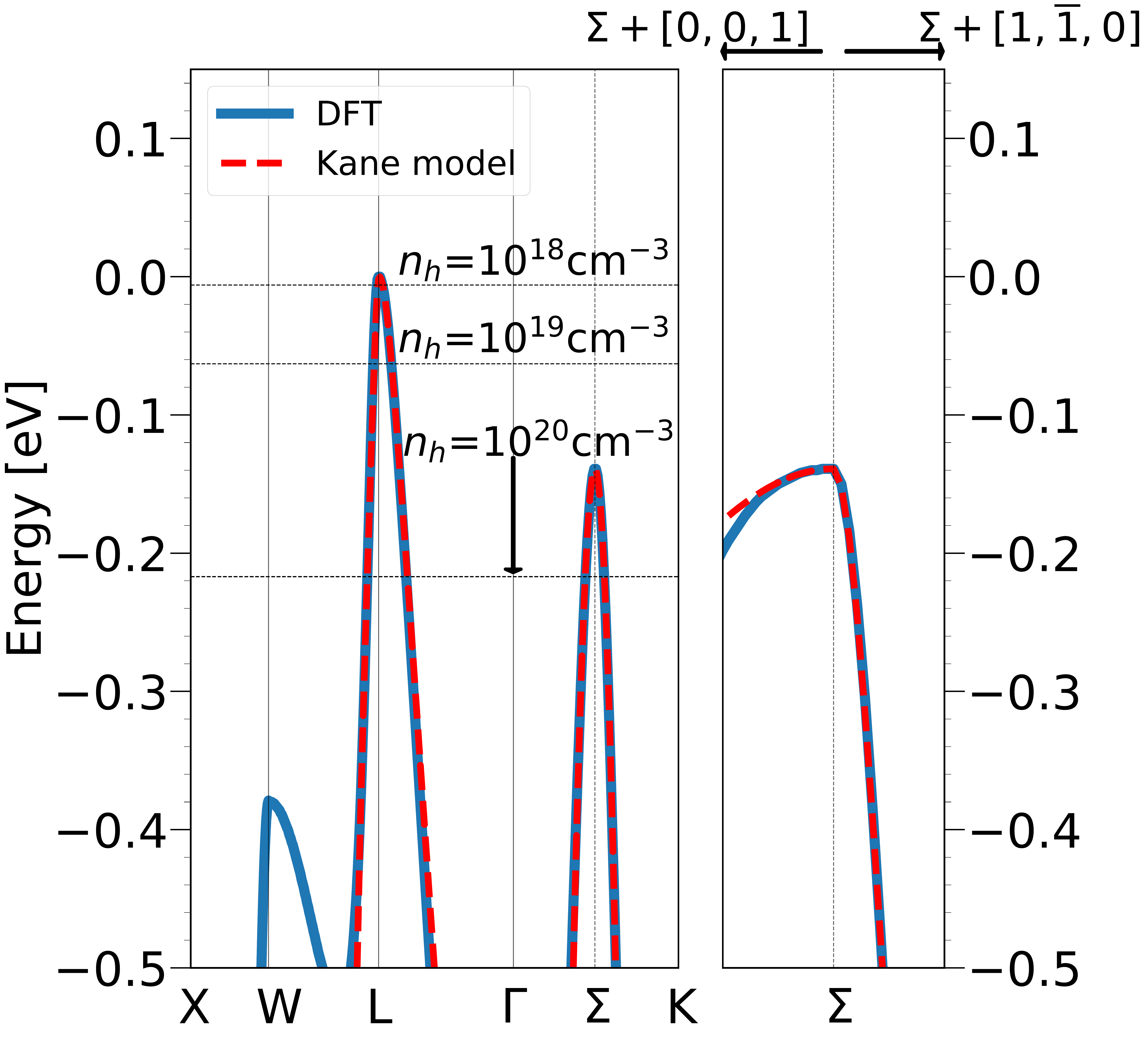}
\end{centering}
\caption{\label{fig:BS} The valence band dispersion of PbTe calculated using density functional theory (DFT, solid blue lines) and the Kane model fitted to DFT (dashed red lines). Horizontal dashed lines refer to the Fermi level at 300 K for the doping concentration of 10$^{18}$ cm$^{-3}$, 10$^{19}$ cm$^{-3}$  and 10$^{20}$ cm$^{-3}$.}
\end{figure}

\begin{table}[!htbp]
\renewcommand{\arraystretch}{1.4}
\begin{threeparttable}
\caption{\label{tab:fp} Parameters used to characterize the electronic bands of $p$-type PbTe computed from first principles: the parallel and perpendicular components of the effective masses at L ($m^L_{\parallel}$ and $m^L_{\perp}$, respectively) and $\Sigma$ ($m^{\Sigma}_{\parallel}$, $m^{\Sigma}_{\perp_{z}}$ and $m^{\Sigma}_{\perp_{xy}}$, respectively), the direct band gaps at L and $\Sigma$ at 0 K ($E^L_g$ and $E^{\Sigma}_g$, respectively) and their temperature coefficients ($\frac{\partial E^L_g}{\partial T}$ and $\frac{\partial E^{\Sigma}_g}{\partial T}$, respectively), and the energy difference between the valence band maxima at L and $\Sigma$ ($\Delta E_{L,\Sigma}$)  at 0 K and 300 K.
$m_e$ stands for free electron mass.}
 \begin{tabular}{c|c|c}
Parameter & Theory & Experiment\\
\hline
$m^{L}_{\parallel}$ ($m_e$) & 0.295 & 0.255$^\text{a}$, 0.310$^\text{b}$ \\
$m^{L}_{\perp}$ ($m_e$) & 0.028 & 0.024$^\text{a}$, 0.022$^\text{b}$\\
$m^{\Sigma}_{\parallel}$ ($m_e$) &  0.179 & \\
$m^{\Sigma}_{\perp_{xy}}$ ($m_e$) & 0.058 & \\
$m^{\Sigma}_{\perp_{z}}$ ($m_e$) & 3.79 & \\
$m^{\Sigma}_{d}$ ($m_e$) & 0.34 & 0.38$^\text{c}$, 0.11-0.45$^\text{d,e}$\\
$E^L_g(T=0$ K) (eV) & 0.237$^\text{f}$ & 0.19$^\text{a,b}$\\
$E^{\Sigma}_g(T=0$ K) (eV) & \color{blue}1.35 & \\
$\frac{\partial E^L_g}{\partial T}$ (eV/K) & 4.4$\times 10^{-4}$ $^\text{g}$  & 3-5.1$\times 10^{-4}$ $^\text{i,j,k,l,m}$ \\
$\Delta E_{L,\Sigma}(T=0$ K) (eV) & 0.14 & 0.14$^\text{d}$, 0.15-0.2$^\text{c,e}$ \\
$\Delta E_{L,\Sigma}(T=300$ K) (eV) & 0.092$^\text{g}$ & 
  \end{tabular} 
\begin{tablenotes}
\item $^\text{a}$ Ref. \cite{pascher2003}
\item $^\text{b}$ Ref. \cite{dalven1974}
\item $^\text{c}$ Ref. \cite{kolomoets68}
\item $^\text{d}$ Ref. \cite{allgaier68}
\item $^\text{e}$ Ref. \cite{svane10}
\item $^\text{f}$ Ref. \cite{murphy18}
\item $^\text{g}$ Ref. \cite{jose19}
\item $^\text{h}$ Ref. \cite{jiang19}
\item $^\text{i}$ Ref. \cite{gibbs13}
\item $^\text{j}$ Ref. \cite{gibson52}
\item $^\text{k}$ Ref. \cite{tauber66}
\item $^\text{l}$ Ref. \cite{saakyan66}
\item $^\text{m}$ Ref. \cite{baleva90}
\end{tablenotes}
\end{threeparttable}
\end{table}

We first discuss the accuracy of our parametrization of the electronic band structure of $p$-type PbTe. In Fig.~\ref{fig:BS} we plot the valence band dispersion obtained using DFT and the Kane model fitted to the DFT results. The left panel of Fig.~\ref{fig:BS} shows the band structure along the selected high symmetry directions in the first BZ. The band dispersion starting from a valence band maximum at $\Sigma$ along the [0,0,1] and [1,-1,0] directions, from which the effective masses $m^{\Sigma}_{\perp_{z}}$ and $m^{\Sigma}_{\perp_{xy}}$ are respectively extracted, is illustrated in the right panel of Fig.~\ref{fig:BS}. The Fermi level values at 300 K that correspond to the doping concentrations of $10^{18}$ cm$^{-3}$, $10^{19}$ cm$^{-3}$ and $10^{20}$ cm$^{-3}$ are shown in dashed horizontal lines\footnote{The Fermi level values change with respect to those presented in Fig. \ref{fig:BS} when the temperature dependence of the L and $\Sigma$ valleys is taken into account.}. It is evident that the Kane model represents the DFT band structure very well for the doping concentrations that are relevant for TE transport\footnote{At larger doping concentrations ($\sim 10^{20}$cm$^{-3}$), the Kane model for the $\Sigma$ valleys deviates from DFT along one of the perpendicular directions ($\Sigma$ + [0,0,1]). This should not significantly affect our results at 300 K since the contribution of the $\Sigma$ valleys is relatively small compared to that of the L valleys even for large doping concentrations. Furthermore, the optimal $ZT$ values for $p$-type PbTe are found at concentrations where the Kane model describes the DFT band structure well ($\sim 4\times 10^{18}$ cm$^{-3}$ at 300~K).}. 
In Table~\ref{tab:fp}, we list all the parameters characterizing the electronic band structure of $p$-type PbTe, which are in very good agreement with available experimental data.

\begin{figure*}[!htbp]
\begin{centering}
\includegraphics[keepaspectratio, width=\textwidth]{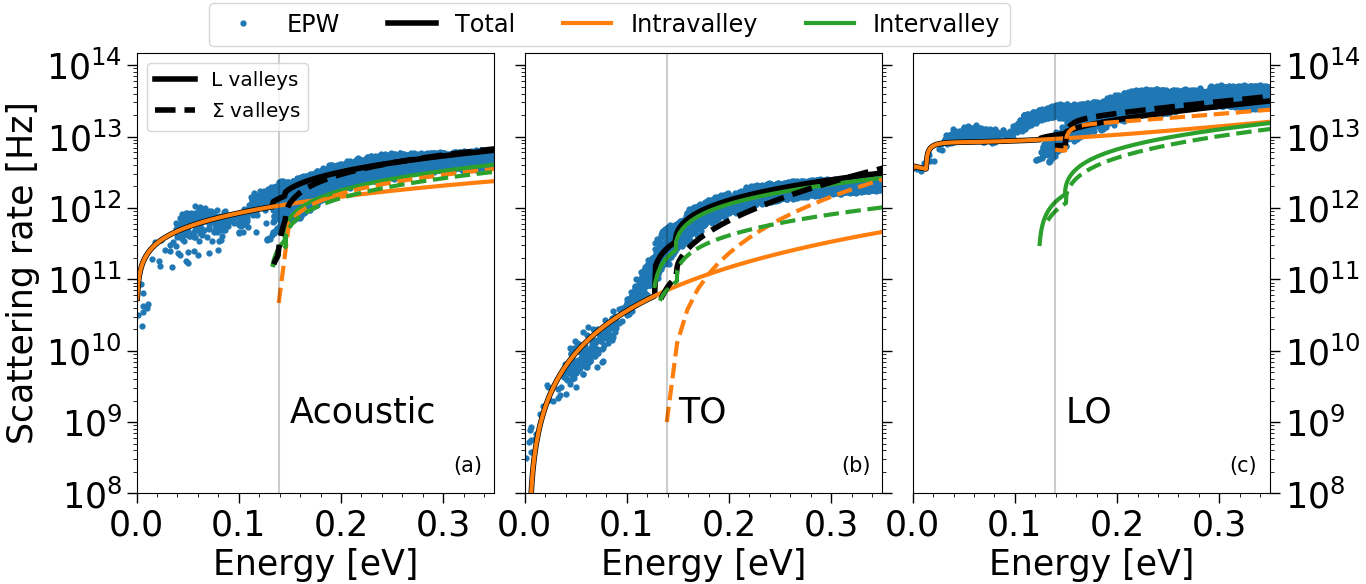}
\end{centering}
\caption{\label{fig:SR} Electron-phonon scattering rates at 300 K for the valence bands of PbTe versus hole energy (from the valence band maxima at L corresponding to zero energy and below). The scattering rates are calculated using the electronic band structure at 0 K and resolved by (a) acoustic, (b) transverse optical (TO) and (c) longitudinal optical (LO) phonon modes. 
Blue dots correspond to the total scattering rates computed using density functional perturbation theory and the electron-phonon Wannier (EPW) approach.
Solid (dashed) lines refer to the scattering rates obtained using our first principles {\color{blue} based} model for the L ($\Sigma$) valleys. The intravalley and intervalley contributions to the total scattering rate (black lines) are given by orange and green lines, respectively. The gray vertical line corresponds to the energy difference between the L and $\Sigma$ valleys at $0$ K.
}
\end{figure*}

\subsection{Electron-phonon scattering rates\label{ssec:sr}}

To validate our model, we compare the scattering rates calculated from the model to those obtained using a general form of the el-ph Hamiltonian (see Eq.~\eqref{eq:me_dfpt} in Appendix~\ref{apdx:DPfDFPT}). 
First, the electron-phonon matrix elements are calculated on coarse $10\times 10 \times 10$ $\bm k$ and $\bm q$ grids using DFPT implemented in the {\sc Quantum espresso} code.
We then use the electron-phonon Wannier interpolation approach \cite{giustino07}, as implemented in the EPW code \cite{ponce16}, to interpolate these matrix elements onto denser $80\times 80 \times 80$ $\bm k$ and $\bm q$ grids using 14 Wannier orbitals for interpolation. 
For the scattering rates obtained from EPW, a broadening parameter of 30 meV was chosen for energy conservation, and screening was not included. We emphasize here that the temperature induced variations of electronic states are not taken into account in the EPW calculations of scattering rates, whose temperature dependence originates only from phonon occupations. In this section, we use the band structure parameters at 0 K to calculate the scattering rates at 300~K without screening, to be consistent with the EPW results. We show in Supplemental Material~\cite{supp} that these scattering rates do not change qualitatively if the temperature variations of the electronic energies are accounted for.

In Fig.~\ref{fig:SR} we plot the scattering rates of the valence bands of PbTe computed from our model and the EPW code. These scattering rates are broken down by phonon modes {\it i.e.} acoustic, transverse optical (TO) and longitudinal optical (LO) modes. Solid (dashed) black lines correspond to the mode resolved scattering rates for the L ($\Sigma$) valleys from our model, while dots refer to the EPW results. The mode dependent scattering rates obtained using our model are in very good agreement with those of our EPW calculations, confirming the validity of the model\footnote{The differences between the scattering rates calculated using our model and the EPW approach near the energies where the $\Sigma$ valleys appear is due to the fact that a broadening parameter of 30~meV was used in the EPW calculations, while energy conservation is exact in our model.}.
\begin{figure*}[!htbp]
\begin{centering}
\includegraphics[keepaspectratio, width=\linewidth]{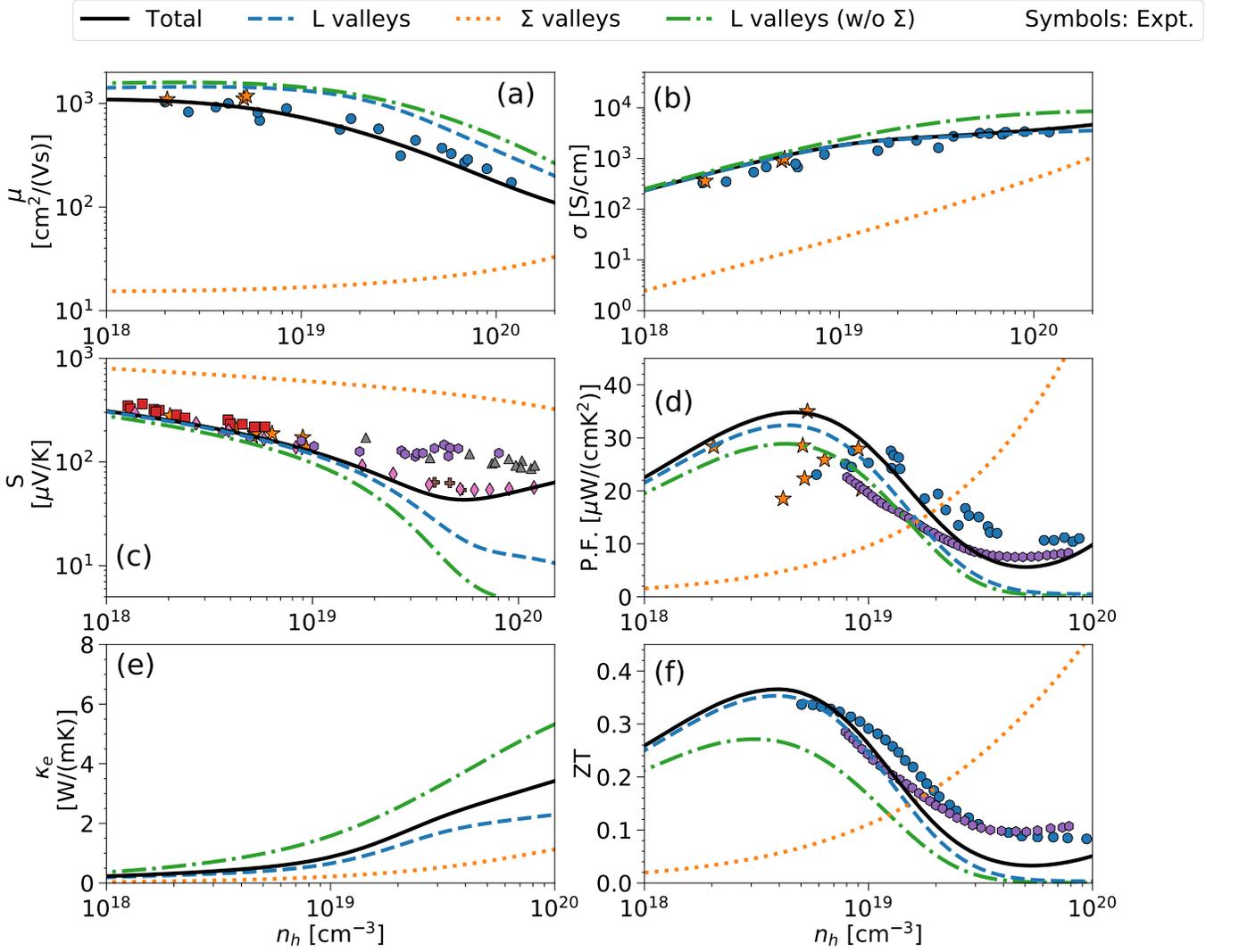}
\end{centering}
\caption{\label{fig:TE} Room temperature thermoelectric transport parameters of $p$-type PbTe:  (a) mobility ($\mu$), (b) conductivity ($\sigma$), (c) Seebeck coefficient ($S$), (d) power factor (PF), (e) electrical thermal conductivity ($\kappa_e$) and (f) figure of merit ($ZT$) as a function of doping concentration $n_h$. Solid black lines show the results obtained using our model and the electronic band structure at 300~K. Dashed blue and dotted orange lines represent the contributions from the L and $\Sigma$ valleys, respectively. Dash-dotted green lines correspond to the contribution from the L valleys in the total absence of the $\Sigma$ valleys (ignoring scattering between the L and $\Sigma$ valleys). Symbols represent measurements: blue circles from Ref. \cite{pei2012}, orange stars from Ref. \cite{vineis08},
purple hexagons from Ref. \cite{nemov98}, red squares from Ref. \cite{rogers71}, magenta diamonds from Ref. \cite{chernik68}, brown plus-signs from Ref. \cite{biswas11}, and gray triangles from Ref. \cite{vinogradova71}.
}
\end{figure*}

Longitudinal optical phonon scattering is the strongest scattering mechanism in $p$-type PbTe at room temperature and low doping concentrations (see Fig.~\ref{fig:SR}), similarly to $n$-type PbTe \cite{jiang18}. This finding is contrary to previous reports \cite{pei14,alalonde11} stating that acoustic modes contribute to the scattering rates of $p$-type PbTe the most. The polar contribution of the LO mode, described by the Fr\"ohlich model, is the most significant contribution to the total LO rates. Nevertheless, the intravalley non-polar and intervalley contributions to the LO scattering rates are not negligible for larger hole energies. 
We show the breakdown of these different contributions to the LO phonon scattering rates in Supplemental Material~\cite{supp}.
 Our results show that acoustic modes are the next dominant source of scattering and transverse optical modes contribute the least to the total scattering rates. These TO phonon modes therefore do not have much influence on electronic transport in $p$-type PbTe, as also seen in our calculations for $n$-type PbTe \cite{jiang18}. On the other hand, the TO modes of PbTe are very soft (with frequencies around 1~THz) and lead to very low lattice thermal conductivity \cite{murphy16,rmurphy17}. Consequently, soft TO modes are one of the key factors responsible for the high thermoelectric figure of merit of both $p$- and $n$-type PbTe.

Next we analyze the intravalley and intervalley contributions to the scattering rates for each type of valleys. For the L valleys, the intravalley and intervalley scattering rates for each phonon mode are given by solid orange and green lines in Fig.~\ref{fig:SR}, respectively. Both intravalley and intervalley scattering are dominated by LO phonons, Fig.~\ref{fig:SR}(c). The intervalley scattering of the L valleys is entirely determined by the scattering between the L and $\Sigma$ valleys in our model. The intravalley contribution to the LO and total scattering rates for the L valleys is larger than the intervalley one for the hole energies up to $\sim 0.1$ eV from the bottom of the $\Sigma$ valleys, Fig.~\ref{fig:SR}(c). This trend is reversed for the larger hole energies, which is a consequence of the large density of states of the $\Sigma$ valleys and sufficiently large el-ph matrix elements between L and $\Sigma$. Most importantly, the scattering rates of the L valleys are increased by a significant amount due to the intervalley scattering, and exhibit a significant change in the energy dependence near the hole energy where the $\Sigma$ valleys appear.

Figure~\ref{fig:SR} also shows the intravalley and intervalley contributions to the phonon mode-resolved scattering rates for the $\Sigma$ valleys (dashed orange and green lines, respectively).  As for the L valleys, LO phonons are the largest contributors to the intravalley and intervalley scattering of the $\Sigma$ valleys, Fig.~\ref{fig:SR}(c). The intervalley scattering for the $\Sigma$ valleys includes both the scattering between the L and $\Sigma$ valleys, and the scattering between different $\Sigma$ valleys. The intervalley scattering of the $\Sigma$ valleys is dominated by the scattering to other $\Sigma$ valleys, as a result of their large density of states. The LO and total intravalley scattering of the $\Sigma$ valleys are larger than the intervalley scattering (see Fig.~\ref{fig:SR}(c)), which can be attributed to the strong Fr\"ohlich and non-polar LO phonon scattering within the $\Sigma$ valleys.

\subsection{Thermoelectric transport parameters}

In this section, we compare our computed room temperature thermoelectric transport parameters of $p$-type PbTe with experiments \cite{li17,pei12,pei2012,vineis08,pei11,nemov98,rogers71,biswas11,chernik68,vinogradova71}.
In Fig.~\ref{fig:TE} we plot the (a) mobility, (b) conductivity, (c) Seebeck coefficient, (d) power factor, (e) electrical thermal conductivity, and (f) figure of merit as a function of doping concentration. Solid black lines indicate our results, while symbols represent the measured data. Here we account for the temperature dependence of the electronic band structure, as well as screening. 
We have also included ionized-impurity scattering \cite{sofo94,ridley99,jiang18}. We find that it has a small impact on the calculated properties because of the large static dielectric constant of PbTe ($\epsilon_s$ = 313.65) \cite{jiang18}.
The computed values of the thermoelectric transport properties agree very well with experiments for a large range of doping concentrations, highlighting the accuracy of our model. In contrast, excluding the temperature dependence of the electronic band structure in the calculation of the thermoelectric transport parameters at 300 K does not lead to such good agreement with experiments (see Supplemental Material \cite{supp}).

To understand the significance of the $\Sigma$ valleys for thermoelectric transport at 300 K, we separate the contributions of the L and $\Sigma$ valleys to the thermoelectric transport parameters, given by blue dashed and orange dotted lines in Fig.~\ref{fig:TE}, respectively. The L valleys contribute mostly to thermoelectric transport, which is expected since the energy difference between the valence band maxima at L and $\Sigma$ at 300 K is still large ($\Delta E_{L,\Sigma} = 0.092$~eV) in comparison to the thermal energy $k_BT$ \cite{jose19}. The contribution from the $\Sigma$ valleys is comparatively small. However, the $\Sigma$ valleys have large density of states that increase both the electrical conductivity and Seebeck coefficient, particularly at high doping concentrations. As a result, the power factor of $p$-type PbTe increases in the presence of the $\Sigma$ valleys.

To analyze this further, green dash-dotted lines in Fig.~\ref{fig:TE} show the contribution of the L valleys to thermoelectric transport in the total absence of the $\Sigma$ valleys i.e. neglecting scattering between the L and $\Sigma$ valleys. We see that including this scattering mechanism (blue dashed lines in Fig.~\ref{fig:TE}) reduces the mobility and conductivity of the L valleys, especially for large doping concentrations. However, the L-$\Sigma$ intervalley scattering increases the Seebeck coefficient of the L valleys at most doping concentrations, resulting in an increase of their power factor. This effect can be understood using the Mott formula $S \sim \left.\frac{1}{\sigma}\frac{\text{d}\sigma}{\text{d}E}\right|_{E=E_F}$ \cite{mehdizadehdehkordi15}, which shows that the Seebeck coefficient is inversely proportional to the conductivity. The intervalley scattering degrades the electronic conductivity, which means that it improves the Seebeck coefficient. Alternatively, the Mott formula can be written as $S \sim \left.\left (\frac{1}{N}\frac{\text{d}N}{\text{d}E}+\frac{1}{v}\frac{\text{d}v}{\text{d}E}+\frac{1}{\tau}\frac{\text{d}\tau}{\text{d}E}\right)\right|_{E=E_F}$, where $N$, $v$ and $\tau$ are the density of states, group velocities and carrier lifetimes \cite{xia19}. The Seebeck coefficient increase due to the L-$\Sigma$ intervalley scattering can be interpreted as a result of an increased energy dependence of the carrier lifetimes of the L valleys in the presence of the L-$\Sigma$ scattering (see Fig.~\ref{fig:SR}). This analysis confirms that, perhaps surprisingly, scattering between the L and $\Sigma$ valleys is not detrimental for the power factor of $p$-type PbTe.

Finally, intervalley scattering between the L and $\Sigma$ valleys reduces the electrical thermal conductivity of the L valleys more strongly than the $\Sigma$ valleys increase the total electrical thermal conductivity, see Fig.~\ref{fig:TE}(e). As a result of the L-$\Sigma$ intervalley scattering, the thermoelectric figure of merit of the L valleys is considerably improved even for low doping concentrations, Fig.~\ref{fig:TE}(f). At high doping, additional transport channels of the $\Sigma$ valleys further increase the total $ZT$. We thus conclude that the $\Sigma$ valleys are indeed beneficial for the thermoelectric performance of $p$-type PbTe at 300~K in a range of doping concentrations. We also note that, even though the energy difference between the L and $\Sigma$ valleys is considerable at 300 K, it is necessary to include the $\Sigma$ valleys in our model to obtain the thermoelectric transport parameters in agreement with experiments.

\section{Conclusion}
We have developed a first principles {\color{blue} based} model of thermoelectric transport in $p$-type PbTe, which includes the temperature induced changes of the L and $\Sigma$ valleys that are responsible for hole conduction. Our calculated thermoelectric transport properties match very well the experimental results at 300 K for different doping levels. We find that longitudinal and transverse optical phonon scattering are the strongest and weakest electron-phonon scattering mechanisms in $p$-type PbTe, respectively. Hole transport at 300 K is mainly determined by the L valleys. Scattering between the L and $\Sigma$ valleys decreases the electrical conductivity and electrical thermal conductivity of the L valleys at 300 K, and increases their Seebeck coefficient, power factor and figure of merit. The $\Sigma$ valleys also provide additional transport channels that increase the total values of all those quantities. These effects arising from the $\Sigma$ valleys lead to the improved thermoelectric figure of merit of $p$-type PbTe at 300 K, suggesting that high valley degeneracy is indeed a desirable trait for efficient thermoelectric materials.

\section{Acknowledgements}
We thank G. Jeffrey Snyder for useful discussions. This project has received funding from the European Union’s Horizon 2020 research and innovation programme under the Marie Skłodowska-Curie grant agreement number 713567. This work is partly supported by Science Foundation Ireland under grant numbers 15/IA/3160 and 13/RC/2077. The later grant is co-funded under the European Regional Development Fund. We acknowledge the Irish Centre for High-End Computing (ICHEC) for the provision of computational facilities.

\appendix
\section{Kane model\label{apdx:kane}}
The energy dispersion of the valence and conduction bands for the L and $\Sigma$ valleys is described by Eqs. \eqref{eq:kane_L} - \eqref{eq:kane}, using the Kane model. The group velocities of e.g. the L valleys are given by \cite{ridley99,jiang18}
\begin{eqnarray}
v_{\parallel(\perp)}^L = \hbar\bigg(\frac{d\gamma_{L}}{dE}\bigg)^{-1}\frac{k^L_{\parallel(\perp)}}{m_{\parallel(\perp)}^L}. 
\end{eqnarray}
The density of states of a valley $v$ ($v=$ L or $\Sigma$) reads \cite{ridley99,jiang18}
\begin{eqnarray}
D_v(E) = \frac{(m_d^v)^{3/2}}{\sqrt{2}\pi^2\hbar^3}\sqrt{\gamma_v(E)}\frac{d\gamma_v(E)}{dE}.
\end{eqnarray}
The overlap integral between the states ${\bm k}$ and ${\bm k'}$ within a valley $v$ has the following form \cite{fawcett70}
\begin{eqnarray}
\begin{aligned}
  &I_v^2(\bm{k},\bm{k'}) = \\ 
  & \hspace{-1em}\frac{(\sqrt{1+\alpha_v E_{\bm k}}\sqrt{1+\alpha_v E_{\bm k'}} + \alpha_v\sqrt{E_{\bm k}E_{\bm k'}}\cos\theta_{\bm k'})^2}{(1+2\alpha_vE_{\bm k})(1+2\alpha_vE_{\bm{ k'}})}.
\end{aligned}
\end{eqnarray} 

We define a new coordinate system by scaling the ${\bm k}$ components along the three principal axes of the ellipsoids so that the energy surfaces are spherical. For the L valleys, this transformation is given as

\begin{eqnarray}\label{eq:k_tran}
\begin{aligned}
k^L_{{\parallel}({\perp})} = \sqrt{\frac{m_{{\parallel}({\perp})}^L}{m_e}}{k}^{*}, 
\end{aligned}
\end{eqnarray}
where $k^*$ represents the transformed wave vectors. The angle that these new wave vectors form with the principal axis is then expressed as \cite{ridley99}
\begin{eqnarray}\label{eq:angle}
\cos\theta_L = \frac{\sqrt{K_m^L}\cos\theta^* + \sin\theta^* (\cos\phi^*+\sin\phi^*)}{\sqrt{1+(K_m^L-1)\cos^2\theta^*}},
\end{eqnarray}
where $K_m^L = \frac{m^L_\parallel}{m^L_\perp}$ is the mass anisotropy coefficient in the L valley. $\theta^*$ and $\phi^*$ are the principal and the azimuthal angles in the new coordinate system, respectively.
Within the $\Sigma$ valleys, the corresponding angle is given as \cite{ridley99}
\begin{equation}\label{eq:angle_sig}
\resizebox{0.5\textwidth}{!}{
$\cos\theta_\Sigma = \frac{\sqrt{m_\parallel}\cos\theta^* + \sqrt{m_{\perp_{xy}}}\sin\theta^*\cos\phi^* + \sqrt{m_{\perp_{z}}}\sin\theta^*\sin\phi^*}{\sqrt{m_\parallel\cos^2\theta^* + m_{\perp_{xy}}\sin^2\theta^*\cos^2\phi^* + m_{\perp_{z}}\sin^2\theta^*\sin^2\phi^* }}$.
}
\end{equation}

\section{Total relaxation times \label{apdx:tsr}}
The total scattering rates and relaxation times are calculated by converting the summations in Eqs.~\eqref{eq:sr2} and \eqref{eq:rt} into integrals via $\sum_{\bm k} \rightarrow \int \frac{V}{(2\pi)^3} d^3{\bm k}$. The relaxation time is then given as
\begin{eqnarray}\label{eq:rt_apdx}
\tau^{-1}({\bm k}) = \frac{V}{(2\pi)^3} \int \sum_{\lambda}W_{\lambda}(\bm{k,k'}) (1-\cos \theta_{\bm k'})\nonumber \\
\times \frac{1-f^0(E_{\bm{k'}})}{1-f^0(E_{\bm{k}})}\delta_{\bm{k\pm q,k'}}  d^3{\bm k'}.
\end{eqnarray}
The Kronecker delta transforms the integral from $\bm{k'}$ space to $\bm q$ space. The coupling term $C_{\lambda,{\bm q}}$ in Eq.~\eqref{eq:el_ph} and the scattering rates $W_{\lambda}(\bm{k,k'})$ are functions of the phonon wave vector ${\bm q}$, and are therefore functions of the angle between ${\bm q}$ and the principal valley axis ($\theta$), and the corresponding azimuthal angle ($\phi$). We represent $\theta$ ($\phi$) as 1000 equally spaced grid points between 0 and $\pi$ ($2\pi$). Rotating the Cartesian coordinate system to align with the principal valley axis as explained in Appendix~\ref{apdx:azimuth}, the final form of Eq.~\eqref{eq:rt_apdx} reads
\begin{eqnarray}\label{eq:rt_apdx2}
\tau^{-1}({\bm k}) &=& \frac{V}{(2\pi)^3}\int_{q_{min}}^{q_{max}} \int_0^{2\pi}\int_{-1}^{1} \sum_{\lambda}W_{\lambda}(\bm{k},q,\theta) \nonumber \\
& {\hspace{-5em}} \times& {\hspace{-3em}} \frac{1-f^0(E_{\bm{k+q}})}{1-f^0(E_{\bm{k}})}(1-\cos \theta_{\bm {k+q}})q^2 dq d(\cos \theta) d\phi.
\end{eqnarray}
We average the electron-phonon matrix elements $|M_{\lambda,{\bm q}}|^2$ in Eq.~\eqref{eq:el_ph} and the scattering rates $W_{\lambda}(\bm{k},q,\theta)=2\pi|M_{\lambda,{\bm q}}|^2/\hbar$ over the azimuthal angles $\phi$ (see Appendix~\ref{apdx:azimuth}). In the Kane model, $q_{min}$ and $q_{max}$ for optical phonons are expressed as \cite{ridley99}
\begin{eqnarray}
q_{min(max)} = k(\chi^+ -(+) 1), \nonumber \\
\chi^+ = 1 +\frac{\gamma(E_{\bm k} + \hbar \omega_{\bm q}) - \gamma(E_{\bm k})}{\gamma(E_{\bm k})},
\end{eqnarray}
for absorption, while for emission
\begin{eqnarray}
q_{min(max)} = k(1-(+)\chi^-), \nonumber \\
\chi^- = 1 - \frac{\gamma(E_{\bm k})-\gamma(E_{\bm k} - \hbar \omega_{\bm q})}{\gamma(E_{\bm k})}.
\end{eqnarray}
For acoustic phonons, $q_{min}$ = 0 and $q_{max} = 2k$ \cite{ridley99}. The integrals and the solution to energy conservation in the Dirac delta function are carried out using the numpy and linear algebra packages in python. Using Eqs. \eqref{eq:kane_L} - \eqref{eq:kane} and \eqref{eq:k_tran}, we transform the wave vector dependent relaxation time $\tau({\bm k})$ to an energy dependent relaxation time $\tau(E)$ for each valley. The energy dependent relaxation times are calculated on 1000 equally spaced grid points between 0 and 10$k_{B}T$. We interpolate those 1000 points to 10$^4$ points within the same energy range for the calculations of the TE transport parameters given in Section~\ref{sec:trans}.

\section{Azimuthal averages of intravalley electron-phonon matrix elements \label{apdx:azimuth}}

\begin{figure}[h]
\begin{centering}
\includegraphics[keepaspectratio, width=9cm]{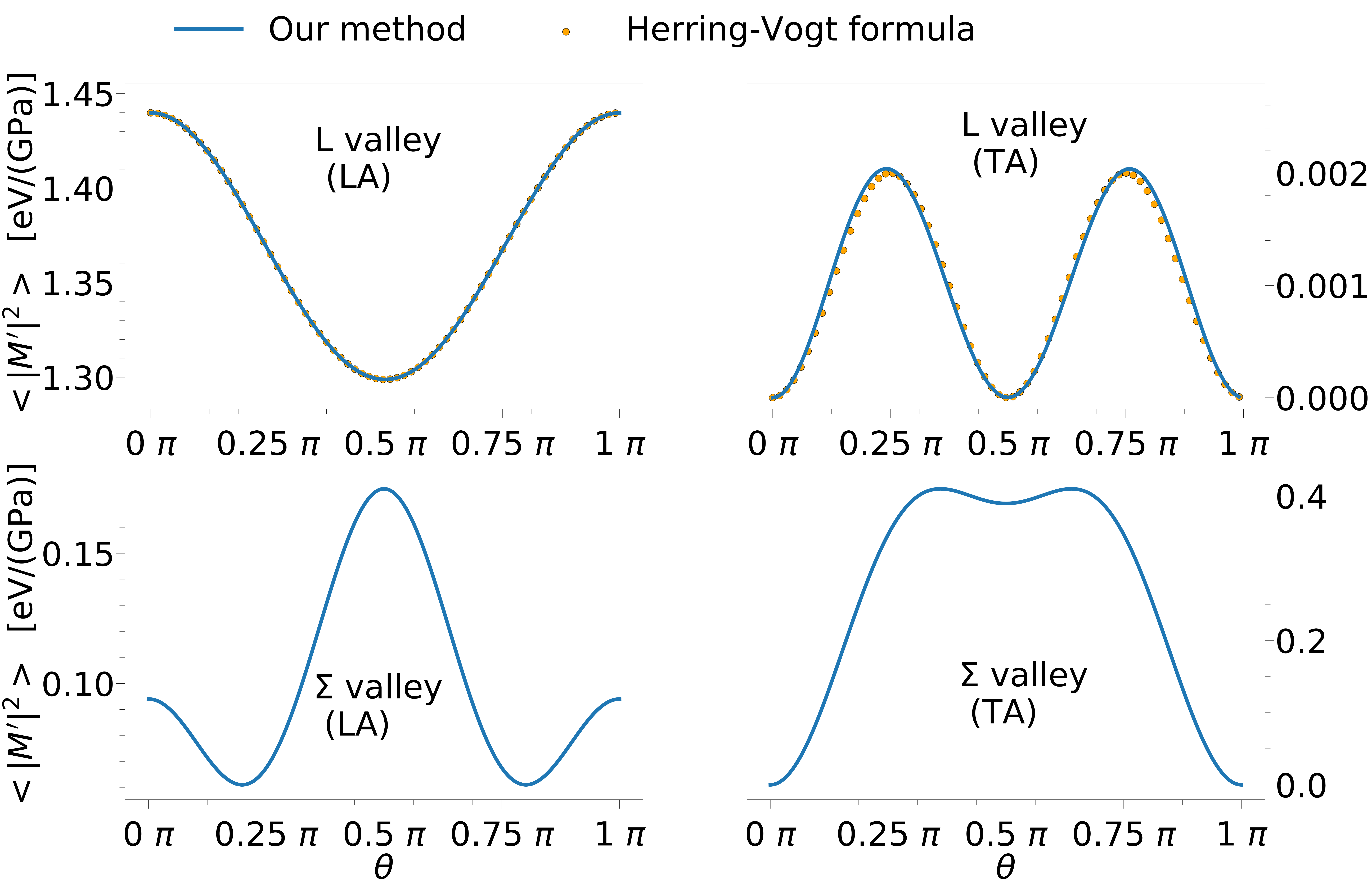}
\end{centering}
\caption{\label{fig:theta} Intravalley acoustic electron-phonon matrix elements (scaled by the factor $\frac{k_BT I^2(\bm{k},\bm{k'})}{V}$  where $I(\bm{k},\bm{k'})$ is the wave function overlap) as a function of the angle between the phonon wave vector and the principal valley axis ($\theta$), averaged over the azimuthal angles, for the L valleys (top panels) and the $\Sigma$ valleys (bottom panels). Left and right panels correspond to longitudinal (LA) and transverse (TA) acoustic modes, respectively. Solid lines show the results obtained our generalized method, while dots refer to a spherical harmonics interpolation technique obtained by Herring and Vogt for the L valley \cite{herring56}.
}
\end{figure}

\begin{figure*}[!htbp]
\begin{centering}
\includegraphics[keepaspectratio, width=\textwidth]{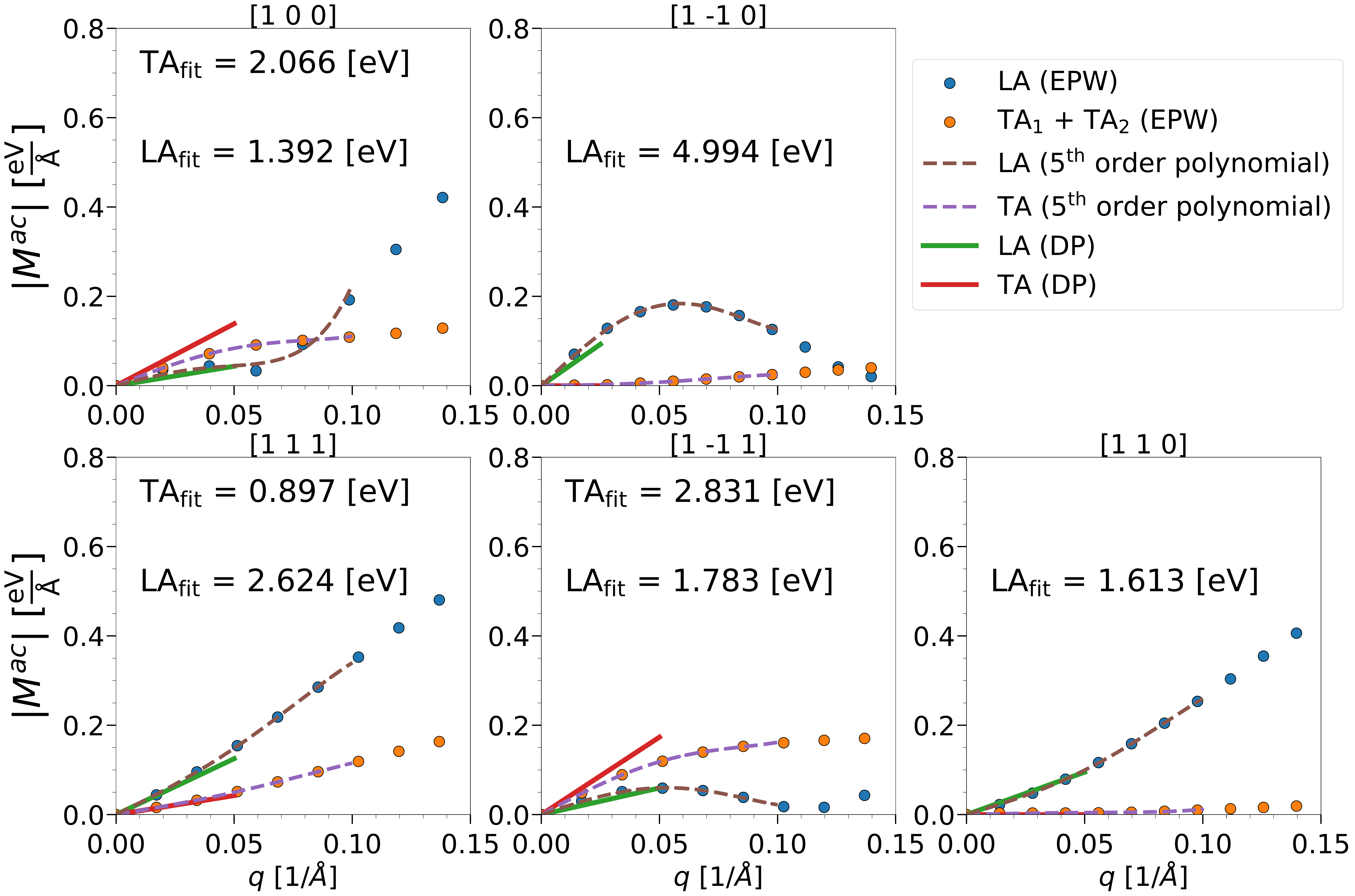}
\end{centering}
\caption{\label{fig:dp_ac} Electron-phonon matrix elements due to acoustic modes between the states $\Sigma$ (0.1875, 0.375, 0.1875) and $\Sigma + {\bm q}$ as a function of the phonon wave vector ${\bm q}$ whose directions are indicated at the top of each panel. Orange and blue dots represent the matrix elements calculated using density functional perturbation theory and electron-phonon Wannier (EPW) approach for transverse and longitudinal acoustic modes, respectively. Dashed lines are the 5$^{\rm th}$ order polynomial fits to the EPW matrix elements. Red (green) lines are obtained using the computed values of deformation potentials (DP) given in Table \ref{tab:dp} for transverse (longitudinal) acoustic modes in the long wavelength limit.
}
\end{figure*}
The el-ph matrix elements and the polarization components given by Eqs.~\eqref{eq:bc} and \eqref{eq:pol}, respectively, depend on the angle between the phonon wave vector ${\bm q}$ and the symmetry axis of the valley (the polar angle $\theta$) and the corresponding azimuthal angles. Averaging over these azimuthal angles, we obtain the matrix elements and the scattering rates that depend only on the polar angles, which are used in Eq.~\eqref{eq:rt_apdx2}. To do this, we need to transform phonon wave vector coordinates with respect to the principal axis into the Cartesian coordinate system. Therefore, the $\bm q$ vector is rotated as follows
\begin{eqnarray}
\begin{aligned}
q_{rot} = Rq,   \\
R = I + [v] + [v]^2\bigg(\frac{1}{1+d}\bigg), \\
\lbrack v \rbrack = 
\begin{pmatrix}
0 &-v_3& v_2 \\
v_3 &0 &-v_1 \\
-v_2& v_1& 0
\end{pmatrix},
\end{aligned}
\end{eqnarray}
where $I$ is the identity matrix, $v_{i}$ are the components of the cross product between (1,0,0) and the valley direction, and $d$ is the cosine of the angle between the valley direction and (1,0,0). Herring and Vogt \cite{herring56} have used a spherical harmonics interpolation technique to get the $\theta$ dependencies of the acoustic matrix elements for L valleys. 
In Fig.~\ref{fig:theta} we plot the scaled matrix elements ($|M^\prime|^2 = \frac{\sum_{ij}\Xi_i \Xi_j f_i f_j}{\sum_{ij} c_{ij} f_i f_j}$) and show that our approach yields almost identical results to those of Herring and Vogt (given in Table V of Ref.~\cite{herring56}), thus confirming the validity of our generalized technique.

\section{Deformation potentials from density functional perturbation theory (DFPT) \label{apdx:DPfDFPT}}

We compute the numerical values of deformation potentials by relating the electron-phonon matrix elements from DFPT to those of our model. In DFPT, the electron-phonon matrix element from a state $\bm{k}$ and band $n$ to a state $\bm{k+q}$ and band $m$ is given as \cite{giustino17}
\begin{eqnarray}\label{eq:me_dfpt}
H^{DFPT}_{mn}({\bm k};\bm{q}\lambda) = \sqrt{\frac{\hbar}{2\omega_{\lambda,\bm{q}}}}\sum_{b,i}\sqrt{\frac{1}{m_b}}e^{\bm{q}\lambda}_{b,i} \nonumber \\
\times \bra{u_{m,\bm{k+q}}} \partial_{b,i,\bm{q}}v^{\rm{ks}} \ket{u_{n,\bm{k}}}, 
\end{eqnarray}
where $e^{\bm{q}\lambda}_{b,i}$ is the $i^{\rm{th}}$ Cartesian component of the phonon eigenvector that corresponds to atom $b$ of mass $m_b$. $\bm q$ and $\lambda$ are the wave vector and branch of the phonon with frequency $\omega_{\lambda,\bm{q}}$ involved in the scattering event. $\partial_{b,i,\bm{q}}v^{\rm{ks}}$ is the lattice periodic part of the first order expansion of the perturbed Kohn-Sham potential \cite{giustino17}. $u_{n,\bm k}$ is the lattice periodic part of the wave function given as $\frac{1}{\sqrt{N_l}}u_{n,\bm k}e^{i\bm{k\cdot r}}$, where $N_l$ is the number of primitive cells.

\section{Intravalley deformation potentials\label{apdx:intra}}

Here we briefly describe our method to compute the numerical values of the acoustic and optical deformation potentials for the L and $\Sigma$ valleys of $p$-type PbTe from first principles, which we have also used to obtain the deformation potentials of the L valleys in $n$-type PbTe \cite{murphy18}. The electron-phonon matrix elements between the states L and L $+$ ${\bm q}$, as well as $\Sigma$ (0.1875, 0.375, 0.1875) and $\Sigma + {\bm q}$, are computed along high-symmetry ${\bm q}$ directions using Eq.~\eqref{eq:me_dfpt} and the EPW code. Dots in Fig.~\ref{fig:dp_ac} show these acoustic matrix elements between $\Sigma$ and $\Sigma + {\bm q}$. Since we are interested in computing the deformation potentials for non-polar phonons, we do not include the Fr\"ohlich contribution in these calculations.

To compute deformation potentials, we find the best multi-directional fit of the electron-phonon matrix elements given by Eq.~\eqref{eq:me_dfpt} to the equations shown in Table \ref{tab:LaTa-cont} for each $\bm q$ direction. Within an L valley, Taylor expansion of the electron-phonon Hamiltonian has odd terms only due to the inversion symmetry of PbTe with respect to its constituent atoms \cite{murphy18}. We have hence fitted the calculated EPW matrix elements with the fifth order polynomials with odd terms only for the L valleys and included even terms for the $\Sigma$ valleys. The numbers in the top left corner of each plot in Fig. \ref{fig:dp_ac} represent the linear terms of the fifth order polynomials for the $\Sigma$ valleys and acoustic phonons. We then fit these linear terms to the expressions in Table \ref{tab:LaTa-cont} for ${\bm q}\rightarrow 0$ using linear regression, and obtain the linearly independent deformation potentials for each valley, given in Table~\ref{tab:dp}. Using those deformation potentials and Table \ref{tab:LaTa-cont}, we finally obtain the electron-phonon matrix elements used in our model, valid for ${\bm q}\rightarrow 0$. They are shown in red and green lines in Fig. \ref{fig:dp_ac} for the $\Sigma$ valleys and acoustic scattering. We note that the zero order non-polar optical matrix element for the $\Sigma$ valleys is not zero (0.0262 eV, see Supplemental Material~\cite{supp}), which was used as the zero order term in the corresponding fifth order polynomials. Finally, the electron-phonon matrix elements and their deformation potential fits for the L valleys and optical modes for the $\Sigma$ valleys are shown in Supplemental Material~\cite{supp}.

\end{document}